\documentclass[a4paper,11pt]{article}

\usepackage{jheppub}
\usepackage[T1]{fontenc}
\usepackage[utf8]{inputenc}
\usepackage{cancel}
\usepackage{amsmath,amssymb,mathtools}
\allowdisplaybreaks[1]
\usepackage{tikz-cd}
\usetikzlibrary{decorations.pathmorphing}
\usepackage{setspace}
\usepackage{float}

\newcommand{\dd}{\mathrm{d}}
\newcommand{\ruh}{r_{\textsc{uh}}}
\newcommand{\aet}{\textnormal{\ae}}
\newcommand{\MD}{M_{\mathrm D}}
\newcommand{\MN}{M_{\mathrm N}}
\newcommand{\MK}{M_{\mathrm K}}

\usepackage[dvipsnames]{xcolor}

\newcommand\Ccancel[2][black]{{\renewcommand\CancelColor{\color{#1}}\cancel{#2}}}
\newcommand{\Ccolor}[2][black]{\textcolor{#1}{#2}}

\title{\boldmath Asymptotic \AE ther Misalignment and Universal-Horizon Mechanics}

\author[a,b,c]{W. Arata,}
\author[d]{M. Ismail,}
\author[a,b,c]{S. Liberati,}
\author[d]{L. Martin,}
\author[d]{D. Mattingly,}
\author[a,b,c]{and G. Neri}

\affiliation[a]{SISSA, via Bonomea 265, 34136 Trieste, Italy}
\affiliation[b]{INFN (Sez.~Trieste), via Valerio 2, 34127 Trieste, Italy}
\affiliation[c]{IFPU, via Beirut 2, 34014 Trieste, Italy}
\affiliation[d]{University of New Hampshire, Durham, NH 03824, USA}

\emailAdd{warata@sissa.it}
\emailAdd{MoustafaKamel.Ismail@unh.edu}
\emailAdd{liberati@sissa.it}
\emailAdd{Luke.Martin@unh.edu}
\emailAdd{David.Mattingly@unh.edu}
\emailAdd{gneri@sissa.it}

\abstract{
Universal horizons trap even arbitrarily fast signals, providing a test
of black-hole thermodynamics in theories with a preferred time direction.
In Einstein--{\ae}ther theory, this direction is defined by a dynamical
field, the {\ae}ther, that can remain non-aligned relative to stationary time at infinity.
In particular, for asymptotically anti-de Sitter black holes, it was so far impossible to recover a first law of black-hole mechanics when such asymptotic tilt of the {\ae}ther is present. 
Here, we resolve this issue for an exact family of static, spherically symmetric solutions. Using the
covariant phase-space formalism, we identify the asymptotic tilt and
the decaying part of the {\ae}ther field as a boundary source and its
conjugate response. Allowing both to vary independently obstructs the
definition of an energy that generates time translations. Fixing either
quantity or imposing a relation between them restores such an energy,
while externally varying the boundary data introduces a work term into the first law. The energies obtained by fixing the source or its response
are related by a Legendre transformation. At the universal horizon,
heat includes both the metric area contribution and an irreducible
{\ae}ther energy flux. The resulting balance separates work at infinity
from horizon heat and explains why the physical temperature cannot
generally be inferred from the coefficient of an area variation in
the black-hole mass.
}

\begin{document}

\maketitle
\raggedbottom

\section{Introduction}
\label{sec:introduction}

Universal horizons provide causal boundaries for excitations whose propagation
speed is unbounded relative to a preferred timelike direction
\cite{BerglundBhattacharyyaMattingly2012,BerglundBhattacharyyaMattingly2013,
CroppEtAl2014}.  Their mechanics is well understood for the one-scale,
asymptotically flat solutions.  The asymptotically anti--de Sitter (AdS)
family, however, contains an additional degree of freedom: the \ae ther need not align
with the stationary Killing field at infinity.  In the exact $c_{14}=0$
solutions, this misalignment produces an extra contribution to the mass
variation and prevents the first law from taking the usual closed
entropy--enthalpy form
\cite{BhattacharyyaMattingly2014,MartinMattingly2024}.

In a previous paper \cite{ArataEtAl2026Elliptic}, we identified the Einstein--\ae ther
counterpart of the elliptic-charge mechanism previously found in
Ho\v{r}ava--Lifshitz gravity~\cite{MartinMattingly2024}.  In the spherically reduced $c_{14}=0$ theory, with the \ae ther denoted by $u^\mu$ and its acceleration
$a^\mu\coloneqq u^\nu\nabla_\nu u^\mu$, where $\nabla_\mu$ is the Levi--Civita connection, and with $f$ a scalar function, we showed that the transformation
\begin{equation}
  \delta u^\mu=f a^\mu,
  \qquad
  \nabla_\mu (f a^\mu)=0,
  \label{eq:elliptic-symmetry}
\end{equation}
is a symmetry with an associated conserved current. We showed that the spatial flux of this current vanishes when the \ae ther is asymptotically aligned, thereby identifying the aligned solutions as a \emph{zero-flux ensemble}.  The charge analysis, however, did not determine the quantity canonically conjugate to a varying misalignment.  For this, one must determine the symplectic structure at the AdS boundary.

This is the purpose of the present work.  We show that the growing and decaying modes $(\alpha,\beta)$ that appear in the solution $(s\cdot\chi) = \alpha r +\beta/r^2$, where $s^\mu$ is a unit vector orthogonal to $u^\mu$ and $\chi^\mu$ is the timelike Killing vector, form a canonical pair.  Allowing both modes to vary produces a \emph{nonzero symplectic flux} through the AdS boundary, so that time translations do not admit a Hamiltonian on the unrestricted solution space. Notably, this is an obstruction that belongs to the phase space itself and its boundary conditions, not to any particular mass formula.  Fixing either mode ($\delta\alpha =0$ or $\delta \beta =0$) or imposing a mixed relation between them (e.g.,~$\beta=\beta(\alpha)$) removes the flux and renders the time-translation charge
integrable. For different boundary conditions, the corresponding Hamiltonians differ by a boundary Legendre transformation.  Alternatively, if the fixed boundary mode is varied externally,
the conjugate response \emph{contributes a work term} and the same identity becomes an open-system first law.

Some care is needed with the notion of energy in this context, as we would like the energy that enters our first law to be a Hamiltonian associated with time translations. The generalized Komar construction gives a finite, background-subtracted state function $\MK$, the Komar mass. By contrast, the covariant phase space (CPS) object associated with time translations is the Iyer--Wald surface one-form $\int_{S^2}\mathbf k_\chi$. Since $\chi^\mu$ generates time translations, an integrable Hamiltonian $H_\chi$ is interpreted as the corresponding energy. However, $\int_{S^2}\mathbf k_\chi$ need not be integrable on an unrestricted solution space. A Hamiltonian is obtained only after the action, its boundary and corner terms, the boundary conditions, and the allowed variations have defined a conserved presymplectic form to which the time translation is tangent \cite{LeeWald1990,IyerWald1994,HarlowWu2020}. Thus, neither radial conservation of the surface one-form $\int_{S^2}\mathbf k_\chi$ nor the existence of the exact state function $\MK$ is sufficient to identify the Komar mass with the Hamiltonian generating time translations.

At the universal horizon, the same logic distinguishes a mechanical decomposition from a thermal first law.  The physical universal-horizon temperature is fixed locally by the peeling rate~\cite{CroppEtAl2014,DelPorroEtAl2023}, while the heat is the complete CPS one-form.  As shown in Ref.~\cite{ArataLiberatiNeri2026}, this one-form contains the metric area contribution together with an irreducible \ae ther Killing-energy flux.  Asymptotic work is a \emph{separate} boundary contribution; it does not alter the local horizon heat.  Thus, a coefficient multiplying $\delta A$ in a chosen parametrization of the Komar mass need not reproduce the local temperature determined by the peeling rate.

The paper is organized as follows. In Sec.~\ref{sec:solution}, we introduce the
theory, the exact AdS solutions, and the relevant variations. In
Sec.~\ref{sec:reduced}, we use the reduced action to identify the boundary
source--response pair. In Sec.~\ref{sec:CPS}, we derive the same result in
the four-dimensional CPS formalism, including the boundary Lagrangian and
corner potential. In Sec.~\ref{sec:ensembles}, we construct the Hamiltonians
associated with different boundary conditions. In Sec.~\ref{sec:open-law},
we derive the universal-horizon balance and the open-system first law. In
Sec.~\ref{sec:aether-symmetry}, we show how the {\ae}ther symmetry of
Ref.~\cite{ArataEtAl2026Elliptic} acts on the boundary data, relate its
Noether flux to the work term of the Neumann ensemble, and connect the CPS response to the \ae ther current
found in Ref.~\cite{ArataEtAl2026Elliptic}. We summarize our results and discuss their implications in
Sec.~\ref{sec:conclusions}. Technical details on the generalized Komar
normalization and on the computation of the {\ae}ther symplectic potential are
collected in Appendices~\ref{app:Komar} and~\ref{app:symplectic-potential},
respectively.

We use metric signature $(-,+,+,+)$.  Bold symbols denote differential forms. The spacetime exterior derivative is $\dd$; the field-space exterior derivative is $\delta$. Greek indices run over $d$ spacetime dimensions, while Latin indices run over the $d-1$ coordinates of a hypersurface, either spacelike as a leaf of the foliation or timelike as the asymptotic boundary.

\section{Theory and exact AdS black holes}
\label{sec:solution}

\subsection{Action and couplings}

Einstein--\ae ther theory supplements the metric $g_{\mu\nu}$ with a dynamical, unit timelike vector field $u^\mu$.  Let $\mathcal M$ denote the spacetime manifold. The bulk action is
\cite{JacobsonMattingly2001}
\begin{equation}
  I_{\mathrm{bulk}} = \frac{1}{16\pi G_{\aet}}\int_{\mathcal M}\boldsymbol\epsilon \left[R - 2 \, \Lambda_0-Z^{\mu\nu}{}_{\rho\sigma}(\nabla_\mu u^\rho)(\nabla_\nu u^\sigma)+\lambda_{\aet}(u^\mu u_\mu + 1)\right],
  \label{eq:action}
\end{equation}
where $G_{\aet}$ is the \ae ther gravitational constant, $\boldsymbol\epsilon = \sqrt{-g}\,\dd^4 x$ is the spacetime volume four-form, $R$ is the Ricci scalar, $\Lambda_0$ is the bare cosmological constant,  $\nabla_\mu$ is the Levi--Civita connection, and $\lambda_{\aet}$ is a Lagrange multiplier enforcing $u^\mu u_\mu=-1$. Moreover,
\begin{equation}
  Z^{\mu\nu}{}_{\rho\sigma}
  \coloneqq c_1g^{\mu\nu}g_{\rho\sigma}
   +c_2\delta^\mu_\rho\delta^\nu_\sigma
   +c_3\delta^\mu_\sigma\delta^\nu_\rho
   -c_4u^\mu u^\nu g_{\rho\sigma},
  \label{eq:Z}
\end{equation}
where $c_i$ are dimensionless couplings. For convenience, we define
\begin{equation}
  \begin{aligned}
    c_{13}&\coloneqq c_1+c_3, & c_{14}&\coloneqq c_1+c_4, \\
    c_{123}&\coloneqq c_1+c_2+c_3, & c_\ell&\coloneqq\frac{2+c_{13}+3c_2}{2}.
  \end{aligned}
  \label{eq:couplings}
\end{equation}
Throughout this work, we restrict to $c_{14}=0$.  In this sector, the \ae ther gravitational constant coincides with the Newton constant measured in the weak-field limit: $G_{\aet}=G_{\mathrm{N}} =: G$.

\subsection{Solution and asymptotic parameters}

We now restrict to the static, spherically symmetric branch of solutions
considered in Refs.~\cite{BhattacharyyaMattingly2014,MartinMattingly2024}.
For the exact $c_{14}=0$ AdS branch, the field equations admit ingoing
Eddington--Finkelstein coordinates with areal radius $r$ in which the metric
can be written as
\begin{equation}
  \dd s^2=-e(r)\,\dd v^2+2\,\dd v\,\dd r
  +r^2\dd\Omega^2,
  \qquad
  \chi= \chi^\mu \partial_\mu = \partial_v.
  \label{eq:EF-metric}
\end{equation}
Here, $v$ is the ingoing Eddington--Finkelstein time coordinate, $e(r)$ is the metric function, $\dd\Omega^2$ is the line element on the unit two-sphere, and $\chi^\mu$ is the stationary Killing vector.  We denote by $s^\mu$ the outward-pointing radial unit vector orthogonal to
$u^\mu$.  The Killing vector can then be decomposed as
\begin{equation}
  \chi^\mu=-(u\cdot\chi)u^\mu+(s\cdot\chi)s^\mu,
  \qquad
  e=(u\cdot\chi)^2-(s\cdot\chi)^2.
  \label{eq:chi-decomposition}
\end{equation}
Within this branch, once the above radial coordinate and normalization have been adopted, the solution is specified by the two radial functions $e(r)$ and $(s\cdot\chi)$. Then $(u\cdot\chi)$ is fixed algebraically by Eq.~\eqref{eq:chi-decomposition}.
The exact $c_{14}=0$ AdS family is characterized by~\cite{BhattacharyyaMattingly2014,MartinMattingly2024}
\begin{align}
  (s\cdot\chi)&=\alpha r+\frac{\beta}{r^2},
  \label{eq:sdotchi-solution}\\
  e(r)&=1-\frac{\Lambda r^2}{3}-\frac{r_0}{r}
        -\frac{c_{13}\,\beta^2}{r^4},
  \label{eq:e-solution}
\end{align}
where
\begin{align}
  \alpha &= \frac{1}{\ell_s},\notag\\
  \beta&=\frac{\ruh^2}{\sqrt{3(1-c_{13})}}
    \sqrt{1+\frac{3\ruh^2}{\ell_u^2}},
  \label{eq:beta-regularity}\\
  r_0&=\frac{4\ruh}{3}+\frac{2\ruh^3}{\ell_u^2}+2\alpha\beta.
  \label{eq:r0-regularity}
\end{align}
Here, $\ell_s$ and $\ell_u$ are parameters that determine the asymptotic behavior of $(s\cdot\chi)$ and $(u\cdot\chi)$, respectively; $\Lambda$ is an effective cosmological constant; and $\ruh$ is the radial location of the universal horizon, which is defined through
\begin{equation}
  \left.(u\cdot\chi)\right|_{r=\ruh}=0.
  \label{eq:UH-def}
\end{equation}

In the language of the two modes $(\alpha,\beta)$, the coefficient $\alpha$ controls the leading asymptotic tilt of
the {\ae}ther relative to $\chi^\mu$, while $\beta$ controls the decaying mode. Note that the latter is fixed by regularity of the {\ae}ther at the universal horizon.

Below, we shall show that $\alpha$ and $\beta$ are, respectively, the boundary source and its response.

The maximally symmetric solution in the same asymptotic branch will be used as
the reference background.  It is obtained by setting $\ruh=0$, and hence
$\beta=r_0=0$, while keeping $\alpha$ and the couplings fixed.

The bare cosmological constant $\Lambda_0$ and the effective cosmological constant $\Lambda$ obey
\begin{equation}
  \frac{1}{\ell_u^2}=c_\ell\alpha^2-\frac{\Lambda_0}{3},
  \qquad
  \frac{\Lambda}{3}=\frac{\Lambda_0}{3}-(c_\ell-1)\alpha^2.
  \label{eq:parameter-relations}
\end{equation}
Varying the second of these relations and using Eq.~\eqref{eq:couplings} gives
\begin{equation}
  \delta\Lambda
  =\delta\Lambda_0-3(c_{13}+3c_2)\alpha\,\delta\alpha.
  \label{eq:delta-Lambda}
\end{equation}
It is important to distinguish variations at fixed action from variations at
fixed effective AdS geometry.  In the first case, $\Lambda_0$ is fixed and
\begin{equation}
  \begin{aligned}
    \delta\Lambda_0&=0, & \delta(\ell_u^{-2})&=2c_\ell\alpha\,\delta\alpha, \\
    \delta\Lambda&=-3(c_{13}+3c_2)\alpha\,\delta\alpha.
  \end{aligned}
  \label{eq:fixed-action-variation}
\end{equation}
Thus, a variation of the asymptotic \ae ther source changes the effective AdS
scale.  If instead the effective geometry is held fixed, one finds
\begin{equation}
  \begin{aligned}
    \delta\Lambda&=0, & \delta(\ell_u^{-2})&=2\alpha\,\delta\alpha, \\
    \delta\Lambda_0&=3(c_{13}+3c_2)\alpha\,\delta\alpha.
  \end{aligned}
  \label{eq:fixed-effective-Lambda-variation}
\end{equation}
The second variation changes a coupling in the action and must therefore be
treated within \emph{extended} thermodynamics. We keep these two variational
problems separate and, whenever the bare cosmological constant is varied, define
\begin{equation}
  P_0\coloneqq-\frac{\Lambda_0}{8\pi G}.
  \label{eq:pressure}
\end{equation}

After subtracting the maximally symmetric reference background, the
generalized Komar construction gives~\cite{MartinMattingly2024}
\begin{equation}
  \MK=\frac{r_0}{2G}
  =\frac{1}{G}\left(
    \frac{2\ruh}{3}+\frac{\ruh^3}{\ell_u^2}+\alpha\beta
  \right).
  \label{eq:Komar-mass}
\end{equation}
The Komar mass is therefore a finite state function on the solution family, so its variation $\delta\MK$ is manifestly exact. The nontrivial question is whether the CPS surface one-form associated with $\chi$ coincides with $\delta\MK$. As we show below, on the unrestricted two-mode solution space, it does not.

\section{Reduced action and the boundary canonical pair}
\label{sec:reduced}
We begin with the spherically reduced \ae ther sector, as it provides the simplest route to identify the candidate boundary source--response pairing. Because variations along the full solution family also induce metric variations, however, this reduced calculation does not automatically establish the symplectic structure of the complete theory. Hence, in Sec.~\ref{sec:CPS}, we perform a full CPS calculation and verify the result.

Per unit Killing time, the $c_{14}=0$ \ae ther sector reduces to
\begin{equation}
  \begin{aligned}
    I_{\aet}^{\mathrm{red}}
    = -\frac{c_{123}}{4G}\int_{r_{\mathrm{in}}}^{r_c}
      \Bigl[\bigl(r(s\cdot\chi)'\bigr)^2+2(s\cdot\chi)^2\Bigr]\,\dd r  -\frac{c_2}{2G}\Bigl[r(s\cdot\chi)^2\Bigr]_{r_{\mathrm{in}}}^{r_c}.
  \end{aligned}
  \label{eq:reduced-action}
\end{equation}
Here, $r_{\mathrm{in}}$ and $r_c$ denote the inner radial endpoint and the outer radial cutoff, respectively.  The endpoints are kept fixed under the radial variation. Upon varying Eq.~\eqref{eq:reduced-action}, we integrate the derivative term in $\delta I_{\aet}^{\mathrm{red}}$ by parts,
\begin{equation}
  \begin{aligned}
    \int_{r_{\mathrm{in}}}^{r_c}r^2(s\cdot\chi)'\,\delta(s\cdot\chi)'\,\dd r
    = \Bigl[r^2(s\cdot\chi)'\,\delta(s\cdot\chi)\Bigr]_{r_{\mathrm{in}}}^{r_c} -\int_{r_{\mathrm{in}}}^{r_c}\bigl[r^2(s\cdot\chi)'\bigr]'
      \delta(s\cdot\chi)\,\dd r,
  \end{aligned}
  \label{eq:reduced-integration-by-parts}
\end{equation}
while the explicit boundary contribution varies as
\begin{equation}
  \delta\!\left[r(s\cdot\chi)^2\right]
  =2r(s\cdot\chi)\,\delta(s\cdot\chi).
  \label{eq:reduced-explicit-endpoint-variation}
\end{equation}
Combining these expressions, we obtain
\begin{equation}
  \begin{aligned}
  \delta I_{\aet}^{\mathrm{red}}
  ={}&\frac{c_{123}}{2G}
    \int_{r_{\mathrm{in}}}^{r_c}
    \left[\big(r^2(s\cdot\chi)'\big)'-2(s\cdot\chi)\right]
    \delta(s\cdot\chi)\,\dd r
  \\
  &-\left[
    \left(\frac{c_{123}}{2G}r^2(s\cdot\chi)'
          +\frac{c_2}{G}r(s\cdot\chi)\right)
    \delta(s\cdot\chi)
  \right]_{r_{\mathrm{in}}}^{r_c}.
  \end{aligned}
  \label{eq:reduced-variation}
\end{equation}
The bulk term yields the radial equation
\begin{equation}
  \big(r^2(s\cdot\chi)'\big)'-2(s\cdot\chi)=0,
  \label{eq:radial-equation}
\end{equation}
whose general solution is precisely Eq.~\eqref{eq:sdotchi-solution}.  

The boundary term in Eq.~\eqref{eq:reduced-variation} identifies the symplectic potential at the outer endpoint
\begin{equation}
  \boldsymbol\Theta_{\aet}^{\mathrm{red}}(r) \coloneqq 
  \frac{-r^2}{2G}\left(c_{123}(s\cdot\chi)' +\frac{2c_2}{r}(s\cdot\chi)\right)
    \delta(s\cdot\chi).
  \label{eq:reduced-potential}
\end{equation}
Taking the field-space exterior derivative and using
\begin{equation}
  \delta(s\cdot\chi)=r\,\delta\alpha+r^{-2}\delta\beta,
  \qquad
  \delta(s\cdot\chi)'=\delta\alpha-2r^{-3}\delta\beta,
  \label{eq:reduced-variations}
\end{equation}
we obtain the boundary presymplectic two-form
\begin{equation}
  \begin{aligned}
    \boldsymbol\Omega_{\aet}^{\mathrm{red}}
    &\coloneqq \delta\boldsymbol\Theta_{\aet}^{\mathrm{red}} \\
    &=-\frac{c_{123}}{2G}r^2\,\delta(s\cdot\chi)'\wedge\delta(s\cdot\chi) \\
    &=-\frac{3c_{123}}{2G}\,\delta\alpha\wedge\delta\beta.
  \end{aligned}
  \label{eq:reduced-symplectic-form}
\end{equation}
Note that this is independent of the radius, as required by radial conservation. Also, for nonzero $c_{123}$, it is nondegenerate on the two-dimensional space of asymptotic modes, identifying $\alpha$ and $\beta$ as a canonical pair within the reduced {\ae}ther sector. The corresponding statement for the full theory will be established independently in Sec.~\ref{sec:CPS}.

For the Dirichlet variational problem, we choose the following minimal local counterterm in the reduced spherical theory to cancel the divergent endpoint variation:
\begin{equation}
  I_{\aet,\mathrm{ct}}^{\mathrm{red}}
  =\frac{c_{13}+3c_2}{4G}\,
    r_c(s\cdot\chi)^2\big|_{r=r_c}.
  \label{eq:reduced-counterterm}
\end{equation}
This choice is not unique, since finite local counterterms may still be added; however, such terms amount to a choice of boundary scheme and do not alter the symplectic
two-form. Indeed, since a boundary Lagrangian contributes to the symplectic potential $\boldsymbol\Theta_{\aet}^{\mathrm{red}}$
through its field-space variation $\delta I_{\aet,\mathrm{ct}}^{\mathrm{red}}$, different choices of counterterm will produce the same boundary presymplectic two-form $\boldsymbol\Omega_{\aet}^{\mathrm{red}} =\delta\boldsymbol\Theta_{\aet}^{\mathrm{red}}$ because $\delta^2 I_{\aet,\mathrm{ct}}^{\mathrm{red}}=0$.

At some finite radius $r_c$, the two endpoint contributions combine to give a ``renormalized'' symplectic potential
\begin{equation}
  \begin{aligned}
 \boldsymbol\Theta_{\aet}^{\mathrm{red,ren}}(r_c) \coloneqq
 &-\frac{1}{2G}\left[
     (c_{13}+3c_2)\alpha r_c^2-\frac{2c_{13}\beta}{r_c}
     \right]\delta(s\cdot\chi)
 \\
 &+\frac{1}{2G}\left[
     (c_{13}+3c_2)\alpha r_c^2
     +\frac{(c_{13}+3c_2)\beta}{r_c}
     \right]\delta(s\cdot\chi).
  \end{aligned}
  \label{eq:reduced-counterterm-cancellation}
\end{equation}
The terms proportional to $r_c^2$ cancel, so that
\begin{equation}
  \boldsymbol\Theta_{\aet}^{\mathrm{red,ren}}(r_c)
  =\frac{3c_{123}}{2G}\,\beta
    \left(\delta\alpha+r_c^{-3}\delta\beta\right).
  \label{eq:reduced-renormalized-potential}
\end{equation}
With this result, we can now identify the work term at the boundary as\footnote{The negative sign is there because the timelike boundary has the opposite orientation to the outer radial endpoint.}
\begin{equation}
  \mathcal W_{\aet}^{\mathrm{ren}}
  \coloneqq -\lim_{r_c\to\infty}
  \boldsymbol\Theta_{\aet}^{\mathrm{red,ren}}(r_c)
  =-\frac{3c_{123}}{2G}\,\beta\,\delta\alpha.
  \label{eq:reduced-source-work}
\end{equation}
This establishes the canonical identification: $\alpha$ is the source and
$(3c_{123}/2G)\beta$ is its response.  The factor of three originates from the
cross term between the $r$ and $r^{-2}$ modes in the symplectic form and is
independent of the normalization chosen for the mass.

This reduced calculation has thus identified the finite pairing in the
\ae ther sector. To establish the corresponding phase-space statement for
the complete theory, and to identify the gravitational energy and the universal-horizon contribution, we must now include the metric sector, the diffeomorphism Noether charge, and the boundary and corner structure of the four-dimensional theory.

\section{Four-dimensional covariant phase space}
\label{sec:CPS}

We now turn to the four-dimensional theory. Our aim is to determine whether the diffeomorphism generated by the stationary Killing vector $\chi^\mu$ admits a Hamiltonian generator on the relevant space of solutions. Because the spacetime has a timelike outer boundary, this question depends not only on the bulk symplectic potential, but also on the boundary and corner terms entering the variational principle.

\subsection{Boundaries, corners, and Hamiltonians}
\label{sec:CPS_boundaries}
Let $\mathcal{M}$ again denote the spacetime manifold, $\Gamma$ its timelike outer boundary, $\mathbf L$ the bulk Lagrangian four-form, and $\boldsymbol\ell$ the boundary Lagrangian three-form. The action takes the form
\begin{equation}
  I=\int_{\mathcal M}\mathbf L+\int_\Gamma\boldsymbol\ell.
  \label{eq:action-with-boundary}
\end{equation}
We denote the dynamical fields collectively by $\phi$. Also, throughout this section, $\dd$ denotes the spacetime exterior derivative, while $\delta$ denotes the exterior derivative on field space. The variation of the bulk Lagrangian can be expressed as
\begin{equation}
  \delta\mathbf L
  =\mathbf E_\phi\,\delta\phi
   +\dd\boldsymbol\Theta(\phi;\delta\phi),
  \label{eq:bulk-variation}
\end{equation}
where $\mathbf E_\phi\doteq0$ are the equations of motion and $\doteq$ denotes equality on shell. The symplectic potential $\boldsymbol\Theta(\phi;\delta\phi)$ is simultaneously a three-form on spacetime and a one-form on field space. Given two field variations $\delta_1\phi$ and $\delta_2\phi$, the associated symplectic current is
\begin{equation}
  \boldsymbol\omega(\phi;\delta_1\phi,\delta_2\phi)
  \coloneqq\delta_1\boldsymbol\Theta(\phi;\delta_2\phi)
   -\delta_2\boldsymbol\Theta(\phi;\delta_1\phi).
  \label{eq:omega-definition}
\end{equation}
Thus, $\boldsymbol\omega$ is a three-form on spacetime and a two-form on field space. For backgrounds satisfying the equations of motion and perturbations satisfying the linearized equations, it provides the local density from which the presymplectic form is constructed.

In the presence of the timelike boundary $\Gamma$, however, the integral of the bulk symplectic current alone does not define the complete phase-space structure.  The pullback of the total boundary variation must also admit a decomposition of the form
\begin{equation}
  \left.\boldsymbol\Theta\right|_\Gamma+\delta\boldsymbol\ell
  =\mathbf B+\dd\boldsymbol\theta.
  \label{eq:boundary-decomposition}
\end{equation}
Here, $\mathbf B$ is the part of the boundary variation that must vanish on variations compatible with the chosen boundary conditions, while $\boldsymbol\theta$ is a two-form on spacetime and a one-form on field space. Because it has support on the intersection of a hypersurface with $\Gamma$, $\boldsymbol\theta$ is referred to as the corner symplectic potential.

\begin{figure}[ht]
    \centering
    \begin{tikzpicture}[x=1cm,y=1cm,scale=0.8]
    
        \fill[Gray!10]
            (2.48,10.75) .. controls (3.20,8.50) and (3.30,7.50) .. (3.30,6.70) .. controls (3.30,5.50) and (2.80,3.50) .. (2.58,1.75) 
            to[bend right=12] (10.00,2.20) .. controls (10.00,3.50) and (9.60,5.50) .. (9.40,7.05) .. controls (9.10,8.00) and (9.00,9.50) .. (9.45,11.80)
            to[bend right=12] (2.48,10.75) -- cycle;
    
        \begin{scope}[shift={(6.29,1.97)}, rotate=3.4]
            \fill[Cerulean!20] (0,0) ellipse (3.7 and 0.75);
            \draw[dashed, line width=0.4pt] (3.7,0) arc[start angle=0, end angle=180, x radius=3.7, y radius=0.75];
            \draw[line width=0.7pt] (-3.7,0) arc[start angle=180, end angle=360, x radius=3.7, y radius=0.75];
        \end{scope}
    
        \begin{scope}[shift={(6.35,6.88)}, rotate=3.3]
            \fill[Cerulean!20] (0,0) ellipse (3.05 and 0.70);
            \draw[dashed, line width=0.4pt] (3.05,0) arc[start angle=0, end angle=180, x radius=3.05, y radius=0.70];
            \draw[line width=0.7pt] (-3.05,0) arc[start angle=180, end angle=360, x radius=3.05, y radius=0.70];
        \end{scope}
    
        \begin{scope}[shift={(5.97,11.28)}, rotate=8.6]
            \fill[Cerulean!20] (0,0) ellipse (3.5 and 0.80);
            \draw[line width=0.7pt] (3.5,0) arc[start angle=0, end angle=180, x radius=3.5, y radius=0.80];
            \draw[line width=0.7pt] (-3.5,0) arc[start angle=180, end angle=360, x radius=3.5, y radius=0.80];
        \end{scope}
    
        \draw[line width=0.8pt]
            (2.48,10.75) .. controls (3.20,8.50) and (3.30,7.50) .. (3.30,6.70)
                     .. controls (3.30,5.50) and (2.80,3.50) .. (2.58,1.75);
        \draw[line width=0.8pt]
            (9.45,11.80) .. controls (9.00,9.50) and (9.10,8.00) .. (9.40,7.05)
                     .. controls (9.60,5.50) and (10.00,3.50) .. (10.00,2.20);

        \draw[->, line width=1.2pt,-stealth] (1.7,0.8) -- (1.7,12.4);
        \node at (0.6,12.2) {\Large $\mathrm{time}$};

        \node at (7.2,11.5) {\Large $\Sigma_+$};
        \node at (7,7)  {\Large $\Sigma$};
        \node at (7.3,2)  {\Large $\Sigma_-$};
    
        \draw[->, line width=1.7pt,-stealth] (9.40,7.05) -- (10.40,7.45);
        \node at (10.75,7.55) {\Large $n$};
        \draw[->, line width=1.7pt,-stealth, ForestGreen] (9.40,7.05) -- (9.6, 8.2);
        \node at (10, 8.2) {\Large $\Ccolor[ForestGreen]{u}$};
        \draw[->, line width=1.7pt,-stealth] (5.3,11.3) -- (5.1,12.8);
        \node at (4.8,12.8) {\Large $t$};
        \draw[->, line width=1.7pt,-stealth] (5.55,7) -- (5.45,8.3);
        \node at (5.15, 8.3) {\Large $t$};
        \draw[->, line width=1.7pt,-stealth] (5.55,2) -- (5.45,3.3);
        \node at (5.1,3.3) {\Large $t$};
    
        \node at (9.5, 9.5) {\Large $\Gamma$};
        \node[anchor=west] at (9.5,12) {\Large $\partial\Sigma_+$};
        \node[anchor=west] at (10,2) {\Large $\partial\Sigma_-$};
    \end{tikzpicture}
    \caption{Representation of the spacetime under consideration. Some parts are highlighted: the lateral boundary $\Gamma$, the two lids $\Sigma_\pm$, the two corners $\partial\Sigma_\pm$, and a generic slice $\Sigma$. The \ae ther $u$ (green) is misaligned with the normal $n$ to the boundary $\Gamma$.}
    \label{fig: spacetime}
\end{figure}
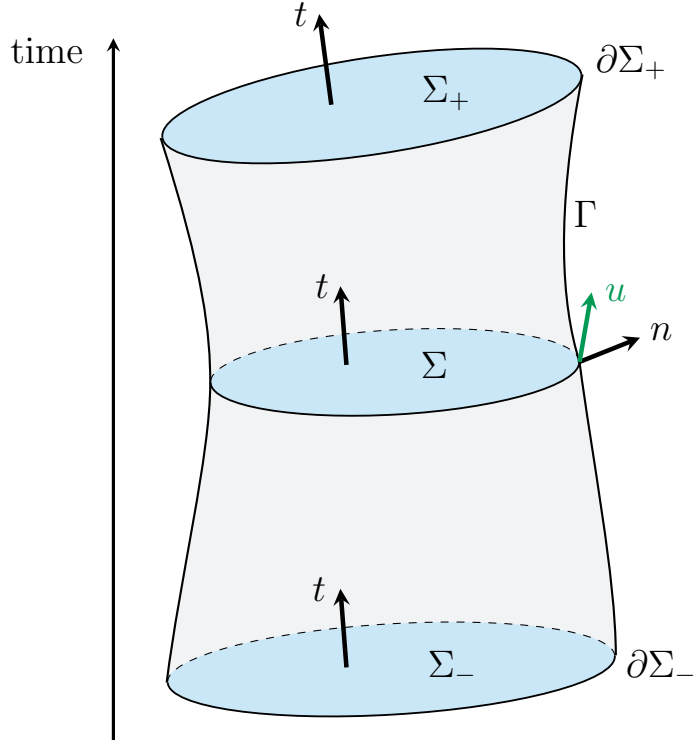

Now, let $\Sigma$ be a hypersurface whose intersection with the timelike boundary is the closed codimension-two surface $\partial\Sigma=\Sigma\cap\Gamma$, as depicted in Fig.~\ref{fig: spacetime}.
The presymplectic form induced by the variational principle is then
\begin{equation}
  \boldsymbol\Omega_\Sigma
  =\delta\left[
    \int_\Sigma\boldsymbol\Theta
    -\int_{\partial\Sigma}\boldsymbol\theta
  \right].
  \label{eq:presymplectic-form}
\end{equation}
Equivalently, its evaluation on two variations is
\begin{equation}
  \boldsymbol\Omega_\Sigma
  (\delta_1\phi,\delta_2\phi)
  =
  \int_\Sigma
  \boldsymbol\omega
  (\phi;\delta_1\phi,\delta_2\phi)
  -
  \int_{\partial\Sigma}
  (\delta\boldsymbol\theta)
  (\delta_1\phi,\delta_2\phi).
  \label{eq:presymplectic-form-expanded}
\end{equation}
The corner term is therefore part of the phase-space structure and must be included together with the bulk symplectic potential.

We next consider a diffeomorphism generated by a field-independent vector field $\xi^\mu$, so that
\begin{equation}
  \delta_\xi\phi=\pounds_\xi\phi,
  \qquad
  \delta\xi=0,
  \label{eq:diffeomorphism-variation}
\end{equation}
where $\pounds_\xi$ is the Lie derivative along $\xi^\mu$. The transformation generated by $\xi^\mu$ is Hamiltonian if there exists a function $H_\xi$ on phase space whose field-space differential is obtained by contracting the presymplectic form with the phase-space vector $\pounds_\xi\phi$.  With the sign convention used here,
\begin{equation}
  -\delta H_\xi
  =
  \pounds_\xi\phi\cdot\boldsymbol\Omega_\Sigma.
  \label{Hamiltonian definition}
\end{equation}
Here, the dot denotes contraction with the first argument of the
field-space two-form $\Omega_\Sigma$. Evaluating both sides of the
preceding identity on an arbitrary allowed variation $\delta\phi$
therefore gives $-(\delta H_\xi)(\delta\phi)=\Omega_\Sigma(\pounds_\xi\phi,\delta\phi)=-\Omega_\Sigma(\delta\phi,\pounds_\xi\phi)$,
where the last equality follows from the antisymmetry of $\Omega_\Sigma$. Cancelling the minus signs and writing
$\delta H_\xi$ for the variation of the Hamiltonian along $\delta\phi$, we obtain
\begin{equation}
  \delta H_\xi
  =
  \boldsymbol\Omega_\Sigma
  (\delta\phi,\pounds_\xi\phi).
  \label{eq:Hamiltonian-variation-definition}
\end{equation}
The problem is therefore to determine whether the right-hand side is an exact one-form on the chosen phase space.

For a diffeomorphism-covariant bulk Lagrangian, define the Noether current three-form
\begin{equation}
  \mathbf J_\xi
  \coloneqq\boldsymbol\Theta(\phi;\pounds_\xi\phi)-\xi\cdot\mathbf L.
  \label{eq:Noether-current}
\end{equation}
Again, $\xi\cdot\mathbf L \coloneqq i_\xi\mathbf L$ denotes contraction of $\xi$ with the first index of the differential form $\mathbf{L}$.  Diffeomorphism covariance implies that $\mathbf J_\xi$ is closed on shell and may locally be written as
\begin{equation}
  \mathbf J_\xi
  \doteq
  \dd\mathbf Q_\xi,
  \label{eq:Noether-charge}
\end{equation}
where $\mathbf Q_\xi$ is the Noether charge two-form.

The relation between the Noether charge and the presymplectic structure follows by varying $\mathbf J_\xi$ in field space while holding the spacetime vector field that generates the diffeomorphism fixed, i.e.,~$\delta \xi =0$. By imposing the background and linearized equations of motion, the variation of Eq.~\eqref{eq:Noether-current} then gives
\begin{equation}
  \begin{aligned}
 \delta\mathbf J_\xi
 &=\delta\boldsymbol\Theta(\phi;\pounds_\xi\phi)
   -\xi\cdot\delta\mathbf L
 \\
 &\doteq \boldsymbol\omega(\phi;\delta\phi,\pounds_\xi\phi)
   +\pounds_\xi\boldsymbol\Theta(\phi;\delta\phi)
   -\xi\cdot\dd\boldsymbol\Theta(\phi;\delta\phi).
  \end{aligned}
  \label{eq:Noether-current-variation}
\end{equation}
Using Cartan's identity for differential forms,
\begin{equation}
  \pounds_\xi\boldsymbol\Theta
  =
  \xi\cdot\dd\boldsymbol\Theta
  +
  \dd\!\left(\xi\cdot\boldsymbol\Theta\right),
  \label{Cartan's Identity}
\end{equation}
this becomes
\begin{equation}
  \delta\mathbf J_\xi
  =
  \boldsymbol\omega
  (\phi;\delta\phi,\pounds_\xi\phi)
  +
  \dd\!\left[
    \xi\cdot\boldsymbol\Theta(\phi;\delta\phi)
  \right].
  \label{eq:Noether-current-variation2}
\end{equation}
On the other hand, varying Eq.~\eqref{eq:Noether-charge} and using $\dd \delta =\delta \dd$, we obtain
\begin{align}\label{eq:Noether-current-variation-charge}
    \delta\mathbf J_\xi\doteq\dd\delta\mathbf Q_\xi.
\end{align}
Comparing Eqs.~\eqref{eq:Noether-current-variation2} and~\eqref{eq:Noether-current-variation-charge}, we obtain the Iyer--Wald identity
\begin{equation}
  \boldsymbol\omega
  (\phi;\delta\phi,\pounds_\xi\phi)
  =
  \dd\mathbf k_\xi,
  \qquad
  \mathbf k_\xi
  \coloneqq
  \delta\mathbf Q_\xi
  -
  \xi\cdot\boldsymbol\Theta(\phi;\delta\phi).
  \label{eq:IW-identity}
\end{equation}
Here, the surface form $\mathbf k_\xi$ is a two-form on spacetime but a one-form on field space. Integrating Eq.~\eqref{eq:IW-identity} over $\Sigma$ gives
\begin{equation}
  \int_\Sigma
  \boldsymbol\omega
  (\phi;\delta\phi,\pounds_\xi\phi)
  =
  \int_{\partial\Sigma}\mathbf k_\xi.
  \label{eq:IW-identity-integrated}
\end{equation}

For the stationary solutions considered here, we take $\xi^\mu=\chi^\mu$, where $\chi^\mu$ is the stationary Killing vector, and consider perturbations tangent to the stationary solution family.  Since
$\pounds_\chi\phi=0$ on the background, Eq.~\eqref{eq:IW-identity} implies\footnote{Although $\pounds_\chi\phi=0$, the surface form itself need not vanish. It contains the variation $\delta\mathbf Q_\chi$ and therefore measures variations between nearby stationary solutions.}
\begin{equation}
  \dd\mathbf k_\chi=0.
  \label{eq:spacetime-closure}
\end{equation}
Consequently, the integral of $\mathbf k_\chi$ over \emph{any} two-sphere is the same:
\begin{equation}
  \int_{S^2_{r_1}}\mathbf k_\chi
  =
  \int_{S^2_{r_2}}\mathbf k_\chi.
  \label{eq:radial-conservation}
\end{equation}
This equation expresses a conservation law on \emph{spacetime}: the surface integral is independent of radius. This does not by itself
imply integrability on \emph{field space}. The integral $\int_{S^2}\mathbf{k}_\chi$ is a one-form on the space of solutions;
for it to be locally expressible as the variation of a state function, it must be closed with respect to the field-space exterior derivative,
\begin{equation}
  \delta\!\int_{S^2}\mathbf k_\chi=0.
  \label{eq:field-space-integrability}
\end{equation}
In other words, while Eq.~\eqref{eq:spacetime-closure} guarantees independence of the radius at which the surface form is evaluated, Eq.~\eqref{eq:field-space-integrability} determines whether the surface one-form can be integrated to a Hamiltonian. The two conditions involve different exterior derivatives, and neither implies the other.

If the bare cosmological constant is varied, the variation is taken in an extended solution space in which the bulk Lagrangian has explicit coupling dependence.  The Iyer--Wald identity is then modified to
\begin{equation}
  \dd\mathbf k_\chi
  =\frac{\delta\Lambda_0}{8\pi G}\,
    \chi\cdot\boldsymbol\epsilon.
  \label{eq:IW-varying-bare-Lambda}
\end{equation}
Hence, $\int_{S_r^2}\mathbf k_\chi$ is not separately independent of radius when $\delta\Lambda_0\neq0$. As we shall show below, adding the corresponding pressure--volume term restores radial conservation on the extended solution family.

We now incorporate the boundary and corner contributions and derive the Hamiltonian associated with $\xi^\mu$. Suppose that $\xi^\mu$ preserves the boundary conditions and that $\mathbf B=0$ for the allowed variations. Evaluating Eq.~\eqref{eq:presymplectic-form-expanded} on $\delta\phi$ and $\pounds_\xi\phi$, and then using Eq.~\eqref{eq:IW-identity} and Stokes' theorem, gives
\begin{align}
  \boldsymbol\Omega_\Sigma
  (\delta\phi,\pounds_\xi\phi)
  =
  \int_{\partial\Sigma}
  \left[
    \delta\mathbf Q_\xi
    -
    \xi\cdot\boldsymbol\Theta(\phi;\delta\phi)
  \right] -
  \int_{\partial\Sigma}
  (\delta\boldsymbol\theta)
  (\delta\phi,\pounds_\xi\phi).
  \label{eq:Omega-boundary-intermediate}
\end{align}
The remaining occurrence of $\boldsymbol\Theta$ can be eliminated using the boundary variational identity.  When $\mathbf B=0$, Eq.~\eqref{eq:boundary-decomposition} becomes
\begin{equation}
  \left.\boldsymbol\Theta(\phi;\delta\phi)\right|_\Gamma
  =
  -\delta\boldsymbol\ell
  +
  \dd\boldsymbol\theta(\phi;\delta\phi).
  \label{eq:Theta-boundary}
\end{equation}
Contracting this equation with $\xi$ and using $\delta\xi=0$ gives
\begin{align}
  -\xi\cdot\boldsymbol\Theta(\phi;\delta\phi)
  &=
  \delta(\xi\cdot\boldsymbol\ell)
  -
  \xi\cdot\dd\boldsymbol\theta(\phi;\delta\phi)
  \notag\\
  &=
  \delta(\xi\cdot\boldsymbol\ell)
  -
  \pounds_\xi\boldsymbol\theta(\phi;\delta\phi)
  +
  \dd\!\left[
    \xi\cdot\boldsymbol\theta(\phi;\delta\phi)
  \right],
  \label{eq:boundary-potential-contracted}
\end{align}
where Cartan's identity was used in the second line.  The final term is exact on the closed corner $\partial\Sigma$ and hence does not contribute to its integral.

It remains to evaluate $\delta\boldsymbol{\theta}$.  Since $\boldsymbol\theta$ is a one-form on field space, its evaluation on the two commuting phase-space vectors $\delta\phi$ and $\pounds_\xi\phi$ is
\begin{equation}
  (\delta\boldsymbol\theta)
  (\delta\phi,\pounds_\xi\phi)
  =
  \delta\!\left[
    \boldsymbol\theta(\phi;\pounds_\xi\phi)
  \right]
  -
  \delta_\xi\!\left[
    \boldsymbol\theta(\phi;\delta\phi)
  \right].
  \label{eq:corner-field-space-derivative}
\end{equation}
For a covariant corner potential $\boldsymbol{\theta}$ and a field-independent $\xi$, its field-space variation along $\delta_\xi\phi=\pounds_\xi\phi$ is the spacetime Lie derivative,
\begin{equation}
  \delta_\xi\!\left[
    \boldsymbol\theta(\phi;\delta\phi)
  \right]
  =
  \pounds_\xi\boldsymbol\theta(\phi;\delta\phi).
  \label{eq:corner-covariance}
\end{equation}
Therefore,
\begin{equation}
  (\delta\boldsymbol\theta)
  (\delta\phi,\pounds_\xi\phi)
  =
  \delta\!\left[
    \boldsymbol\theta(\phi;\pounds_\xi\phi)
  \right]
  -
  \pounds_\xi\boldsymbol\theta(\phi;\delta\phi).
  \label{eq:corner-field-space-identity}
\end{equation}
Substituting Eqs.~\eqref{eq:boundary-potential-contracted} and
\eqref{eq:corner-field-space-identity} into
Eq.~\eqref{eq:Omega-boundary-intermediate}, the terms involving
$\pounds_\xi\boldsymbol\theta(\phi;\delta\phi)$ cancel.  We obtain
\begin{equation}
  \begin{aligned}
  \boldsymbol\Omega_\Sigma
  (\delta\phi,\pounds_\xi\phi)
  =
  \delta\int_{\partial\Sigma}
  \Bigl[
    \mathbf Q_\xi
    +
    \xi\cdot\boldsymbol\ell
    -
    \boldsymbol\theta(\phi;\pounds_\xi\phi)
  \Bigr].
  \end{aligned}
  \label{eq:Hamiltonian-derivation}
\end{equation}
Comparison with Eq.~\eqref{eq:Hamiltonian-variation-definition} identifies the Hamiltonian generator as
\cite{HarlowWu2020,ArataLiberatiNeri2026}
\begin{equation}
  H_\xi
  =
  \int_{\partial\Sigma}
  \Bigl[
    \mathbf Q_\xi
    +
    \xi\cdot\boldsymbol\ell
    -
    \boldsymbol\theta(\phi;\pounds_\xi\phi)
  \Bigr]
  +\text{constant}.
  \label{eq:Hamiltonian-definition}
\end{equation}
We emphasize that this result is conditional on the existence of a boundary phase space for which $\mathbf B=0$, $\xi$ preserves the boundary conditions, and the presymplectic form is finite and conserved. In particular, if Eq.~\eqref{eq:field-space-integrability} is not satisfied, then $H_\xi$ does not exist; in such a case, additional boundary conditions or boundary degrees of freedom will be required.

For the stationary family considered below,
$\pounds_\chi\phi=0$, and hence the explicit term
$\boldsymbol\theta(\phi;\pounds_\chi\phi)$ in
Eq.~\eqref{eq:Hamiltonian-definition} vanishes.  Nevertheless,
$\boldsymbol\theta$ remains essential through its contribution to the
presymplectic form in Eq.~\eqref{eq:presymplectic-form}.  In the present
theory, the \ae ther kinetic term generates no independent corner potential,
so that $\boldsymbol\theta_{\aet}=0$, whereas the metric sector carries the
usual Einstein--Hilbert corner contribution
\cite{ArataLiberatiNeri2026}.  The boundary Lagrangian contains the
Gibbons--Hawking--York term together with the appropriate AdS subtraction.
Finite boundary terms may further implement a Legendre transformation that
exchanges which member of a boundary canonical pair is held fixed.  Such a
transformation changes both the Hamiltonian and the admissible variations and
will give rise below to the different boundary ensembles.

The remaining task is therefore to evaluate the radially conserved
Iyer--Wald surface one-form and determine whether it is closed on the solution space under consideration.  We begin by calculating the asymptotic
symplectic potential.

\subsection{Asymptotic symplectic potential}

To evaluate the asymptotic contribution, it is convenient to work in static coordinates. While the ingoing Eddington--Finkelstein chart used in Sec.~\ref{sec:solution} was well suited to describing the black-hole solution globally because of its regularity, static coordinates make the treatment of constant-$r$ surfaces as the timelike boundary particularly simple for practical calculations. In this chart, we have
\begin{equation}
  \dd s^2=-e(r)\,\dd t^2+\frac{\dd r^2}{e(r)}
  +r^2\dd\Omega^2,
  \qquad
  \chi=\partial_t.
  \label{eq:static-metric}
\end{equation}
The orthonormal vectors in the $(t,r)$ plane can be written directly in terms
of their contractions with the Killing vector:
\begin{align}
 u_\mu\dd x^\mu
 &= (u\cdot\chi)\,\dd t-\frac{(s\cdot\chi)}{e}\,\dd r,
 &
 u^\mu\partial_\mu
 &= -\frac{(u\cdot\chi)}{e}\,\partial_t-(s\cdot\chi)\,\partial_r,
 \label{eq:aether-static-components}
 \\
 s_\mu\dd x^\mu
 &= (s\cdot\chi)\,\dd t-\frac{(u\cdot\chi)}{e}\,\dd r,
 &
 s^\mu\partial_\mu
 &= -\frac{(s\cdot\chi)}{e}\,\partial_t-(u\cdot\chi)\,\partial_r.
 \label{eq:radial-static-components}
\end{align}
Let $n^\mu$ be the outward-pointing unit normal to a surface of constant $r$, and
choose the orientation
$\boldsymbol\epsilon_\Gamma=n\cdot\boldsymbol\epsilon$, where $\boldsymbol\epsilon$ is again the spacetime volume form.  Then
\begin{equation}
  n_\mu\dd x^\mu=\frac{\dd r}{\sqrt e},
  \qquad
  \left(n\cdot u\right) =-\frac{(s\cdot\chi)}{\sqrt e}.
  \label{eq:outer-unit-normal}
\end{equation}
For $\Lambda<0$, Eq.~\eqref{eq:outer-unit-normal} gives
\begin{equation}
  \lim_{r\to\infty} (n\cdot u) = -\frac{\alpha}{\sqrt{-\Lambda/3}}.
  \label{eq:asymptotic-aether-tilt}
\end{equation}
This explains why, in the AdS solution above, we identified the linear coefficient in $(s\cdot\chi)$ with the misalignment; in other words, $\alpha=1/\ell_s$ parametrizes the leading asymptotic tilt.

We now move forward with calculating the {\ae}ther symplectic potential. We first define
\begin{equation}
  Y^\mu{}_\rho\coloneqq Z^{\mu\nu}{}_{\rho\sigma}\nabla_\nu u^\sigma.
  \label{eq:Y-tensor}
\end{equation}
With $q_{\mu\nu}\coloneqq g_{\mu\nu}-n_\mu n_\nu$ denoting the induced metric, the pullback of the {\ae}ther symplectic potential to the boundary is given by
\cite{Foster2006,ArataLiberatiNeri2026}
\begin{equation}
  \begin{aligned}
    \left.\boldsymbol\Theta_{\aet}\right|_\Gamma
    =-\frac{\boldsymbol\epsilon_\Gamma}{16\pi G}\Bigl[n_\mu\bigl(u^\mu Y^{ij}-2u^iY^{(\mu j)}\bigr)\delta q_{ij} +2n_\mu Y^{\mu i}\delta u_i +2n_\mu n_\nu Y^{\mu\nu}\delta(n\cdot u)\Bigr],
  \end{aligned}
  \label{eq:aether-potential-boundary}
\end{equation}
where we use $Y^{\mu\nu}\coloneqq Y^\mu{}_{\rho}g^{\rho\nu}$ and parentheses for symmetrization, $Y^{(\mu\nu)}\coloneqq \tfrac12(Y^{\mu\nu}+Y^{\nu\mu})$. Keeping the cutoff radius fixed, the relevant variations are
\begin{equation}
  \begin{aligned}
    \delta q_{tt}&=-\delta e,\\
    \delta u_t&=\delta(u\cdot\chi), \\
    \delta u_r&=-\frac{\delta(s\cdot\chi)}{e}
       +\frac{(s\cdot\chi)}{e^2}\delta e, \\
    \delta(n\cdot u)&=-\frac{\delta(s\cdot\chi)}{\sqrt e}
       +\frac{(s\cdot\chi)}{2e^{3/2}}\delta e.
  \end{aligned}
  \label{eq:cutoff-variations}
\end{equation}
Upon substitution into Eq.~\eqref{eq:aether-potential-boundary}, the terms
proportional to $\delta e$ and $\delta(u\cdot\chi)$ cancel. The integral over a sphere of areal radius $r$ reduces to (see Appendix~\ref{app:symplectic-potential} for details)
\begin{equation}
  \int_{S_r^2}\chi\cdot\boldsymbol\Theta_{\aet}
  =\frac{r^2}{2G}
   \left[c_{123}(s\cdot\chi)'
        +\frac{2c_2}{r}(s\cdot\chi)\right]
   \delta(s\cdot\chi).
  \label{eq:aether-potential-sphere}
\end{equation}
Up to an overall sign that depends on the orientation choice, this expression is precisely the four-dimensional counterpart of Eq.~\eqref{eq:reduced-potential}.

On the exact family in Eqs.~\eqref{eq:sdotchi-solution}--\eqref{eq:r0-regularity}, we have
\begin{equation}
  \begin{aligned}
    \int_{S_r^2}\chi\cdot\boldsymbol\Theta_{\aet}
    =\frac{1}{2G}\Bigl[(c_{13}+3c_2)\alpha r^3\delta\alpha
       +(c_{13}+3c_2)\alpha\delta\beta -2c_{13}\beta\delta\alpha
       -2c_{13}\beta r^{-3}\delta\beta\Bigr].
  \end{aligned}
  \label{eq:aether-potential-expanded}
\end{equation}
The first term grows as $r^3$ and is present in the unrenormalized symplectic
potential.  At fixed action, i.e., $\delta\Lambda_0=0$, it cancels when the variation of the Noether charge is combined with the metric contribution in the full surface
form, as we shall see in Eq.~\eqref{eq:surface-form-exact-family}. If instead $\alpha$ is varied while the effective cosmological constant $\Lambda$ is
held fixed, the metric variation no longer cancels the growing term.  Within
the static spherical sector, the required counterterm is
\begin{equation}
  \int_{S_r^2}\chi\cdot\boldsymbol\ell_{\aet}^{\mathrm{ct}}
  =-\frac{c_{13}+3c_2}{4G}\,r(s\cdot\chi)^2.
  \label{eq:four-dimensional-counterterm}
\end{equation}
This renders the boundary potential finite within the sector under
consideration.\footnote{It should be noted that Eq.~\eqref{eq:four-dimensional-counterterm} does not, by
itself, determine a unique covariant counterterm beyond spherical symmetry.}
Moreover, its variation is exact on field space and therefore leaves the
symplectic form unchanged.  But since
$\delta\Lambda_0\neq0$ in this case, it is an extended variational problem distinct from
the fixed-action phase space.

A final remark concerns the sign of the counterterm in Eq.~\eqref{eq:four-dimensional-counterterm}, which is opposite to that of the reduced counterterm in Eq.~\eqref{eq:reduced-counterterm}. As for the symplectic potentials in Eqs.~\eqref{eq:reduced-potential} and~\eqref{eq:aether-potential-sphere},
this reflects the orientation of the timelike boundary, which is opposite to that of the outer radial endpoint in the reduced action. The \emph{relative} sign between the counterterm and the symplectic potential, which is what matters for the cancellation of the growing term, is the same in both cases.

\subsection{Surface-charge one-form}

At fixed coordinate radius, we define the surface-charge one-form
\begin{equation}
  \not\!\delta\mathcal H_\chi(r)
  \coloneqq\int_{S_r^2}\mathbf k_\chi
  =\int_{S_r^2}\left(
    \delta\mathbf Q_\chi-\chi\cdot\boldsymbol\Theta
  \right).
  \label{eq:surface-one-form-definition}
\end{equation}
The slashed variation emphasizes that this quantity is a one-form on solution
space and is not assumed to be integrable \emph{a priori}.  A function
$\mathcal H_\chi$ can be introduced only after integrability has been
established on the chosen phase space. In the Einstein--Hilbert sector,
\begin{equation}
  \int_{S_r^2}\mathbf Q_{\chi,\mathrm{GR}}
  =\frac{r^2e'}{4G},
  \qquad
  \int_{S_r^2}\mathbf k_{\chi,\mathrm{GR}}
  =-\frac{r}{2G}\delta e.
  \label{eq:metric-charge}
\end{equation}
The metric Komar integral is twice the first expression in
Eq.~\eqref{eq:metric-charge}, reflecting the standard normalization difference
between the Komar integral and the Iyer--Wald Noether charge.

The \ae ther contribution to the integrated Noether charge is
\begin{equation}
  \int_{S_r^2}\mathbf Q_{\chi,\aet}
  =(s\cdot\chi)\,
   \frac{r^2}{2G}\left[
     c_{123}(s\cdot\chi)'
     +\frac{2c_2}{r}(s\cdot\chi)
   \right].
  \label{eq:aether-Noether-charge}
\end{equation}
Varying Eq.~\eqref{eq:aether-Noether-charge} and subtracting the symplectic
potential in Eq.~\eqref{eq:aether-potential-sphere}, we find
\begin{equation}
  \int_{S_r^2}\mathbf k_{\chi,\aet}
  =(s\cdot\chi)\,\delta\!\left\{
    \frac{r^2}{2G}\left[c_{123}(s\cdot\chi)'
      +\frac{2c_2}{r}(s\cdot\chi)\right]\right\}.
  \label{eq:aether-surface-form}
\end{equation}

Now evaluate the two sectors on the exact solution family.  The metric
contribution is
\begin{equation}
 -\frac{r}{2G}\delta e
 =\frac{r^3}{6G}\delta\Lambda
  +\frac{\delta r_0}{2G}
  +\frac{c_{13}\beta}{Gr^3}\delta\beta,
 \label{eq:metric-surface-form-evaluation}
\end{equation}
while the {\ae}ther contribution becomes
\begin{equation}
  \begin{aligned}
    \int_{S_r^2}\mathbf k_{\chi,\aet}
    =\frac{1}{2G}\Bigl[(c_{13}+3c_2)\alpha r^3\delta\alpha
       +(c_{13}+3c_2)\beta\delta\alpha -2c_{13}\alpha\delta\beta
       -2c_{13}\beta r^{-3}\delta\beta\Bigr].
  \end{aligned}
  \label{eq:aether-surface-form-evaluation}
\end{equation}
Using Eq.~\eqref{eq:delta-Lambda}, the terms proportional to
$r^3\alpha\delta\alpha$ cancel between the metric and \ae ther sectors, and
the same occurs for the $r^{-3}\beta\delta\beta$ terms.  We are therefore left
with
\begin{equation}
  \not\!\delta\mathcal H_\chi(r)
  =\delta\MK
   +\frac{c_{13}+3c_2}{2G}\beta\,\delta\alpha
   -\frac{c_{13}}{G}\alpha\,\delta\beta
   +\frac{r^3}{6G}\delta\Lambda_0\,,
  \label{eq:surface-form-exact-family}
\end{equation}
where the $r^3$ term is produced by the variation of the bare
cosmological constant.  

Let us now use the definition in Eq.~\eqref{eq:pressure} and add the corresponding pressure--volume contribution $\frac{4\pi r^3}{3}\,\delta P_0=-\frac{r^3}{6G}\,\delta\Lambda_0$ to both sides of Eq.~\eqref{eq:surface-form-exact-family}. We obtain
\begin{equation}
  \not\!\delta\mathcal H_\chi(r)
  +\frac{4\pi r^3}{3}\,\delta P_0
  =\delta\MK
   +\frac{c_{13}+3c_2}{2G}\beta\,\delta\alpha
   -\frac{c_{13}}{G}\alpha\,\delta\beta\,.
  \label{eq:radially-conserved-one-form}
\end{equation}
The pressure--volume term cancels the explicit radial dependence of the surface-charge one-form, so the complete combination on the left-hand side is independent of $r$.

We now take the field-space exterior derivative of Eq.~\eqref{eq:radially-conserved-one-form}, keeping the coordinate radius $r$, Newton's constant $G$, and the couplings $c_i$ fixed. Since $M_K$ is a state function, $\delta^2 M_K=0$. The remaining terms are evaluated using the Leibniz rule and the nilpotence of $\delta$. In particular, $ \delta(\beta\,\delta\alpha) =\delta\beta\wedge\delta\alpha$ and $\delta(\alpha\,\delta\beta) =\delta\alpha\wedge\delta\beta$, so one finds
\begin{equation}
\begin{aligned}
\delta\!\left[
\not\!\delta \mathcal H_\chi(r)
+\frac{4\pi r^3}{3}\,\delta P_0
\right]
&=
\frac{c_{13}+3c_2}{2G}\,
\delta\beta\wedge\delta\alpha
-\frac{c_{13}}{G}\,
\delta\alpha\wedge\delta\beta
\\
&=
-\frac{3c_{123}}{2G}\,
\delta\alpha\wedge\delta\beta.
\end{aligned}
\end{equation}
Here, the second equality follows from the antisymmetry of the wedge product. At fixed coordinate radius,
the pressure term is itself closed,
\begin{equation}
    \delta\!\left(\frac{4\pi r^3}{3}\,\delta P_0\right)=0\,,
\end{equation}
even when $\delta P_0\neq0$, so that in the end the field-space exterior derivative of the left-hand side is
\begin{equation}
  \delta \left[
  \not\!\delta\mathcal H_\chi(r) \right]
  =-\frac{3c_{123}}{2G}\,
    \delta\alpha\wedge\delta\beta,
  \label{eq:integrability-obstruction}
\end{equation}
and we see that both the coefficient and wedge structure in Eq.~\eqref{eq:integrability-obstruction} agree with the reduced two-form in
Eq.~\eqref{eq:reduced-symplectic-form}.

For nonzero $c_{123}$, Eq.~\eqref{eq:integrability-obstruction} shows that the surface-charge one-form is not closed when both asymptotic modes vary
independently.\footnote{This does not contradict $\delta^2=0$: the notation $\not\!\delta\mathcal H_\chi$ denotes a one-form that is not assumed to be the
differential of a state function.} Time-translation charges are therefore integrable only on an admissible submanifold of the solution space, namely one on which the two-form in Eq.~\eqref{eq:integrability-obstruction} pulls back to zero. Since this two-form is proportional to $\delta\alpha\wedge\delta\beta$,
any condition that restricts $(\alpha,\beta)$ to a curve suffices: for example, one may choose $\delta\alpha=0$ (Dirichlet boundary conditions), $\delta\beta=0$ (Neumann
boundary conditions),\footnote{This terminology reflects the usual distinction between fixing a boundary field value and its conjugate normal momentum: Dirichlet conditions fix the source $\alpha$, while Neumann-type conditions fix the response $(3c_{123}/2G)\beta$, which plays the role of the renormalized
momentum conjugate to that source.} or a mixed relation $\beta=\beta(\alpha)$. The corresponding Hamiltonians are constructed in Sec.~\ref{sec:ensembles}.

\section{Boundary conditions and Hamiltonians}
\label{sec:ensembles}

Equation~\eqref{eq:integrability-obstruction} shows that time translations
admit no Hamiltonian on the unrestricted two-mode solution space.  To obtain
an integrable charge, we must impose a boundary condition and pull back the
surface form to the corresponding phase space.  We carry this out within the
stationary spherical sector; extending the construction beyond this sector
requires the complete boundary variational problem described around
Eq.~\eqref{eq:Hamiltonian-definition}.

\subsection{Dirichlet boundary conditions: \texorpdfstring{$\delta\alpha=0$}{}} 
Let us start by defining 
\begin{equation}
  \MD\coloneqq\MK-\frac{c_{13}}{G}\alpha\beta.
  \label{eq:Dirichlet-energy}
\end{equation}
Substitution into Eq.~\eqref{eq:radially-conserved-one-form} gives
\begin{equation}
  \not\!\delta\mathcal H_\chi(r)
  +\frac{4\pi r^3}{3}\,\delta P_0
  =\delta\MD+\frac{3c_{123}}{2G}\beta\,\delta\alpha.
  \label{eq:Dirichlet-decomposition}
\end{equation}
On the Dirichlet phase space, the last term vanishes and the Hamiltonian is
\begin{align}
    H_\chi^{\mathrm D}=\MD,
\end{align}
up to the choice of reference energy. This identification refers to the phase space at fixed bare action, $\delta P_0=0$. If the bare cosmological constant
is also varied, Eq.~\eqref{eq:Dirichlet-decomposition} shows that $\delta\MD$ equals the radially conserved combination that includes the pressure--volume
term, so that $\MD$ plays the role of an enthalpy in the sense of extended black-hole thermodynamics~\cite{KastorRayTraschen2009}; the same remark applies to $\MN$ below. If the source $\alpha$ is instead changed externally, the same equation identifies $(3c_{123}/2G)\beta$ as its response and the final term as
the work supplied through the boundary.

\subsection{Neumann boundary conditions: \texorpdfstring{$\delta\beta=0$}{}} 
In this case we define
\begin{equation}
  \MN\coloneqq\MK+\frac{c_{13}+3c_2}{2G}\alpha\beta.
  \label{eq:Neumann-energy}
\end{equation}
The conserved one-form becomes
\begin{equation}
  \not\!\delta\mathcal H_\chi(r)
  +\frac{4\pi r^3}{3}\,\delta P_0
  =\delta\MN-\frac{3c_{123}}{2G}\alpha\,\delta\beta.
  \label{eq:Neumann-decomposition}
\end{equation}
Thus, on the Neumann phase space, the Hamiltonian is
\begin{align}
   H_\chi^{\mathrm N}=\MN. 
\end{align}
The two ensemble
energies are related by
\begin{equation}
  \MN-\MD=\frac{3c_{123}}{2G}\alpha\beta,
  \label{eq:Legendre-transform}
\end{equation}
which is the boundary Legendre transformation that exchanges source and
response.

\subsection{Mixed boundary conditions: \texorpdfstring{$\beta=\beta(\alpha)$}{}} 
A mixed boundary condition restricts the two modes to a curve
$\beta=\beta(\alpha)$.  The pullback of the surface form to this curve is
integrable and yields
\begin{equation}
  H_\chi^{\mathrm{mix}}
  =\MD+\frac{3c_{123}}{2G}
    \int \beta(\alpha)\,\dd\alpha.
  \label{eq:mixed-Hamiltonian}
\end{equation}

Thus, within the two-mode sector, fixing either mode or imposing a mixed
relation between them produces an integrable Hamiltonian.  A different
possibility would be to enlarge the system by adding boundary degrees of
freedom whose symplectic form cancels
Eq.~\eqref{eq:integrability-obstruction}.  Such an extension lies beyond the
phase spaces considered here.

\section{Universal-horizon balance and the open-system first law}
\label{sec:open-law}

We can now transport the conserved surface form from infinity to the universal
horizon.  The variation is first evaluated on a sphere of fixed coordinate
radius and the result is then restricted to $r=\ruh$.  Since the combined
one-form in Eq.~\eqref{eq:radially-conserved-one-form} is independent of $r$,
varying the location of the final evaluation surface produces no additional
term.

We denote the complete surface-charge one-form at the universal horizon by
\begin{equation}
  \not\!\delta\mathcal H_\chi^{\textsc{uh}}
  \coloneqq\left.\int_{S_r^2}\mathbf k_\chi\right|_{r=\ruh}.
  \label{eq:horizon-charge-one-form}
\end{equation}
Evaluating Eq.~\eqref{eq:radially-conserved-one-form} at $r=\ruh$ and using Eq.~\eqref{eq:horizon-charge-one-form}, then solving for the Komar mass variation, gives the open-system first
law on the unrestricted solution family,
\begin{equation}
  \delta\MK
  =\not\!\delta\mathcal H_\chi^{\textsc{uh}}
   +\frac{4\pi\ruh^3}{3}\,\delta P_0
   -\frac{c_{13}+3c_2}{2G}\beta\,\delta\alpha
   +\frac{c_{13}}{G}\alpha\,\delta\beta.
  \label{eq:Komar-open-first-law}
\end{equation}
The final two terms describe the work done at the asymptotic boundary.  Applying
the field-space exterior derivative $\delta$ to their sum gives
$+(3c_{123}/2G)\delta\alpha\wedge\delta\beta$, which cancels the field-space
exterior derivative of the horizon charge, as required by
Eq.~\eqref{eq:Komar-open-first-law}.  The nonintegrable horizon contribution is therefore
balanced by the nonintegrable asymptotic work on the unrestricted family.
This is the balance law for an open system.

Rewriting Eq.~\eqref{eq:radially-conserved-one-form} at $r=\ruh$ in terms of the Dirichlet and
Neumann Hamiltonians gives, respectively,
\begin{align}
  \delta\MD
  &=\not\!\delta\mathcal H_\chi^{\textsc{uh}}
   +\frac{4\pi\ruh^3}{3}\,\delta P_0
   -\frac{3c_{123}}{2G}\beta\,\delta\alpha,
  \label{eq:Dirichlet-first-law}\\
  \delta\MN
  &=\not\!\delta\mathcal H_\chi^{\textsc{uh}}
   +\frac{4\pi\ruh^3}{3}\,\delta P_0
   +\frac{3c_{123}}{2G}\alpha\,\delta\beta.
  \label{eq:Neumann-first-law}
\end{align}
When $\delta\alpha=0$ or $\delta\beta=0$, the final term vanishes and these
expressions become the closed Hamiltonian first laws in their respective
ensembles.  If the boundary source or response is changed externally, the
same term is interpreted as the work exchanged with the boundary.

Next, the physical temperature of the universal horizon is determined by the
peeling surface gravity~\cite{CroppEtAl2014,DelPorroEtAl2023},
\begin{equation}
  T_{\textsc{uh}}=\frac{\kappa_{\textsc{uh}}}{\pi},
  \qquad
  \kappa_{\textsc{uh}}
  \coloneqq\frac12\left|\chi\cdot a\right|_{r=\ruh}.
  \label{eq:UH-temperature}
\end{equation}
The local CPS analysis of Ref.~\cite{ArataLiberatiNeri2026} gives the
complete horizon heat,
\begin{equation}
  \not\!\delta\mathcal H_\chi^{\textsc{uh}}
  =T_{\textsc{uh}}\,\delta\left(\frac{A}{4G}\right)
   +\not\!\delta Q_{\aet}^{\mathrm{heat}},
  \qquad
  A=4\pi\ruh^2.
  \label{eq:complete-horizon-heat}
\end{equation}
The second term is the contribution from the \ae ther Killing-energy flux through the horizon.  If we let $T_{\aet}^{\mu\nu}$ denote the \ae ther stress tensor, $H$ the horizon segment, $\boldsymbol\epsilon_\mu$ the directed hypersurface volume three-form, and $\widehat\chi^\mu$ the horizon generator normalized to unit surface gravity before the infinite-speed limit is taken, then, in the finite-speed disformal construction~\cite{ArataLiberatiNeri2026}, the flux takes the form
\begin{equation}
  \not\!\delta Q_{\aet}^{\mathrm{heat}}
  =\frac{\kappa_{\textsc{uh}}}{8\pi G}\,
   \delta\!\int_H
   T_{\aet}{}^\mu{}_\nu\,\widehat\chi^{\nu}
   \boldsymbol\epsilon_\mu.
  \label{eq:aether-heat}
\end{equation}
This result is then extended to the universal horizon by taking an infinite-speed probe limit~\cite{ArataLiberatiNeri2026}.

Substituting Eq.~\eqref{eq:complete-horizon-heat} into the two ensemble laws, Eqs.~\eqref{eq:Dirichlet-first-law} and~\eqref{eq:Neumann-first-law}, we obtain their thermal form
\begin{align}
  \delta\MD
  &=T_{\textsc{uh}}\,\delta\left(\frac{A}{4G}\right)
   +\not\!\delta Q_{\aet}^{\mathrm{heat}} +\frac{4\pi\ruh^3}{3}\,\delta P_0 -\frac{3c_{123}}{2G}\beta\,\delta\alpha,  \label{eq:Dirichlet-thermal-first-law}\\
  \delta\MN
  &=T_{\textsc{uh}}\,\delta\left(\frac{A}{4G}\right)
   +\not\!\delta Q_{\aet}^{\mathrm{heat}} +\frac{4\pi\ruh^3}{3}\,\delta P_0 +\frac{3c_{123}}{2G}\alpha\,\delta\beta.
  \label{eq:Neumann-thermal-first-law}
\end{align}

\section{{\AE}ther symmetry and the boundary ensembles}
\label{sec:aether-symmetry}

In our previous paper~\cite{ArataEtAl2026Elliptic}, we analyzed the symmetry in
Eq.~\eqref{eq:elliptic-symmetry} in the static, spherically symmetric sector
with $c_{14}=0$. The infinitesimal transformation takes the form
\begin{equation}
  u_\mu \mapsto u_\mu + f a_\mu,
  \label{eq:aether_symmetry}
\end{equation}
so that $\delta u^\mu = f a^\mu = f(a\cdot s)s^\mu$. Since $a_\mu$ is orthogonal to
$u_\mu$, this transformation preserves the unit-norm constraint at linear
order, and it is a symmetry provided that the scalar function $f$ satisfies
$\nabla_\mu(fa^\mu)=0$~\cite{ArataEtAl2026Elliptic}. In the static spherical
sector, this condition takes the manifestly elliptic form
\begin{equation}
  \vec{\nabla}_i (f a^i) + f a^2 = 0,
  \label{eq:elliptic-f}
\end{equation}
where $\vec{\nabla}_i$ is the covariant derivative intrinsic to the preferred
spatial leaf and $a^2\coloneqq a_i a^i$. For the solution in
Eqs.~\eqref{eq:sdotchi-solution}--\eqref{eq:r0-regularity}, the allowed
functions $f$ form a one-parameter family labeled by a constant
$\mathcal{C}$~\cite{ArataEtAl2026Elliptic},
\begin{equation}
  f(r) = \frac{\mathcal{C}\,r^3}{(r^3 - \ruh^3)(3r^3+ 2 \ell_u^2 \, \ruh+6\ruh^3)}.
  \label{eq:f-solution}
\end{equation}

Requiring the orthonormality relations $u^2=-1$, $s^2=1$, and $(u\cdot s)=0$
to be preserved, the transformation extends to $s^\mu$ as
\begin{equation}
  \delta_f u^\mu=\eta\, s^\mu, \qquad
  \delta_f s^\mu=\eta\, u^\mu, \qquad
  \eta\coloneqq f\,(a\cdot s),
  \label{eq:aether-boost}
\end{equation}
where $\delta_f$ denotes the variation generated by the symmetry. This is an
$r$-dependent $\mathrm{SO}(1,1)$ boost in the Lorentzian $u$--$s$ plane. Since
the Killing vector is not transformed, $\delta_f(s\cdot\chi)=\eta\,(u\cdot\chi)$,
and Eq.~\eqref{eq:f-solution} gives
$\delta_f(s\cdot\chi)=-\mathcal{C}/(3\ell_u^2 r^2)$. The functional form of
$(s\cdot\chi)$ in Eq.~\eqref{eq:sdotchi-solution} is therefore preserved, and
the asymptotic data transform as
\begin{equation}
  \alpha\mapsto\alpha, \qquad \beta\mapsto\beta-\frac{\mathcal{C}}{3\ell_u^2}.
  \label{eq:alpha-beta-transformation}
\end{equation}
Equivalently, $\delta_f\alpha=0$ and
\begin{equation}
  \delta_f\beta = r^2 f\,(a\cdot s)(u\cdot\chi) = -\frac{\mathcal{C}}{3\ell_u^2},
  \label{eq:deltaf-beta}
\end{equation}
which is independent of $r$ as a consequence of the elliptic
condition Eq.~\eqref{eq:elliptic-f}.

The associated spatial flux through the asymptotic timelike boundary
is~\cite{ArataEtAl2026Elliptic}
\begin{equation}
  \mathcal{F}=(u\cdot\chi)\,r^2 f\,\vartheta\,(a\cdot s)
  =\frac{\alpha\,\mathcal{C}}{\ell_u^2},
  \label{eq:flux-elliptic}
\end{equation}
where $\vartheta\coloneqq\nabla_\mu u^\mu$ is the {\ae}ther expansion. On the
solution family, 
\begin{align}
    \vartheta=-r^{-2}\big[r^2(s\cdot\chi)\big]'=-3\alpha,
\end{align}
so that, comparing with Eq.~\eqref{eq:deltaf-beta},
\begin{equation}
  \mathcal{F}=\vartheta\,\delta_f\beta=-3\alpha\,\delta_f\beta.
  \label{eq:flux-beta}
\end{equation}
The misaligned sector thus carries a nonvanishing boundary flux associated with
the {\ae}ther symmetry, whereas the flux vanishes in the aligned limit.

The Noether flux of the {\ae}ther symmetry, as normalized in
Ref.~\cite{ArataEtAl2026Elliptic}, carries the additional factor
$-8\pi c_{123}$ arising from the current and the angular integration,
$\mathcal{F}_{\aet}=-8\pi c_{123}\,\mathcal{F}$.  Restoring the overall factor
$1/(16\pi G)$ of the action, we find
\begin{equation}
  \frac{\mathcal{F}_{\aet}}{16\pi G}=\frac{3c_{123}}{2G}\,\alpha\,\delta_f\beta.
  \label{eq:work-flux}
\end{equation}
The right-hand side is the Neumann work term of
Eq.~\eqref{eq:Neumann-thermal-first-law} evaluated along the symmetry
direction. Hence, \emph{when the response is shifted by the {\ae}ther symmetry at
fixed source, the work exchanged with the boundary coincides with the Noether
flux of the symmetry}. Since both sides of Eq.~\eqref{eq:work-flux} are
proportional to $\delta_f\beta$, this identity does not depend on the
normalization of $\mathcal{C}$.

The symmetry thus acts only on the response, leaving the source fixed. By
itself, this does not select a preferred ensemble. It does, however, make the
Neumann form of the first law the convenient one for following the symmetry:
when $\beta$ is varied externally, as the symmetry does, the corresponding work
appears explicitly in Eq.~\eqref{eq:Neumann-thermal-first-law} and is given by
Eq.~\eqref{eq:work-flux}. In the aligned limit $\alpha\to0$, both the
flux Eq.~\eqref{eq:flux-elliptic} and this work term vanish, and the first law
reduces to the aligned universal-horizon law of
Ref.~\cite{ArataLiberatiNeri2026}.

\subsection{Relation to the \ae ther charge}
\label{sec:paper-I}

In Ref.~\cite{ArataEtAl2026Elliptic}, we derived the current associated with
Eq.~\eqref{eq:elliptic-symmetry} and showed that its spatial flux $\mathcal{F}$ is controlled by the asymptotic misalignment $\alpha = 1/\ell_s$. Both the flux and the elliptic \ae ther charge
vanish in the aligned limit, $\alpha=0$.  The CPS analysis identifies this same
sector as the zero-source Dirichlet phase space.

The two constructions capture complementary aspects of the boundary physics. The elliptic current arises from a symmetry of the reduced {\ae}ther action
and measures charge transport along a preferred leaf, whereas the CPS response follows from the variation of the full gravitational action and is canonically conjugate to the source $\alpha$. The two quantities need not coincide numerically, but their relation is nonetheless precise. In Ref.~\cite{ArataEtAl2026Elliptic}, we showed that alignment removes the elliptic flux, while the present analysis shows that the two boundary modes of the
misaligned sector carry a nonvanishing symplectic flux, as shown in Eqs.~\eqref{eq:reduced-symplectic-form} and~\eqref{eq:integrability-obstruction}. This fixes the normalization of the associated work term. Moreover, along the direction generated by the {\ae}ther symmetry, this work term coincides with the Noether flux of the symmetry, Eq.~\eqref{eq:work-flux}.

\section{Conclusions}
\label{sec:conclusions}

This work was motivated by an open problem in the mechanics of universal horizons in asymptotically AdS spacetimes. For the exact $c_{14}=0$ black-hole solutions
of Einstein--{\ae}ther theory, the {\ae}ther can be tilted with respect to the stationary Killing vector at infinity, and this misalignment, controlled by the
coefficient $\alpha$ of the growing mode of $(s\cdot\chi)$, had so far prevented the formulation of a first law~\cite{BhattacharyyaMattingly2014,MartinMattingly2024}. In a previous paper~\cite{ArataEtAl2026Elliptic}, we identified a symmetry of the {\ae}ther sector whose flux vanishes precisely for aligned solutions, but left undetermined the quantity conjugate to a varying misalignment. Here, we have identified this quantity from the symplectic structure at the AdS boundary and derived the corresponding first law.

Our main results are threefold. First, the growing and decaying modes in $(s\cdot\chi)=\alpha r+\beta/r^2$ form a canonical pair, with $\alpha$ the boundary source and $(3c_{123}/2G)\beta$ its response. The full CPS analysis shows that the Iyer--Wald surface one-form is finite and radially conserved, but that its field-space exterior derivative equals the same symplectic form already found in the reduced action. Hence, when both modes vary, time translations admit no Hamiltonian. It is worth stressing that this obstruction lies in the phase space structure, not in a particular mass formula, and that the Komar mass, although a state function, is not the generator of time translations. 

Second, integrability is restored by fixing the source (Dirichlet), fixing the response (Neumann), or imposing a relation between them. These choices yield, respectively, the Hamiltonians in Eqs.~\eqref{eq:Dirichlet-energy},~\eqref{eq:Neumann-energy}, and~\eqref{eq:mixed-Hamiltonian}. The first two choices are related by the boundary Legendre transformation in Eq.~\eqref{eq:Legendre-transform}.

Third, transporting the conserved surface form to the universal horizon gives the open-system first laws Eqs.~\eqref{eq:Dirichlet-thermal-first-law}
and~\eqref{eq:Neumann-thermal-first-law}. The temperature is the one fixed by the peeling surface gravity $\kappa_\textsc{uh}$. The horizon heat contains, besides the area term, the {\ae}ther Killing-energy flux identified in Ref.~\cite{ArataLiberatiNeri2026}, and the misalignment enters only through a work term at infinity. In the limit $\alpha\to0$, these laws reduce to the aligned case law of Ref.~\cite{ArataLiberatiNeri2026}.

Several lessons follow from this analysis. The first concerns the new work terms. In the Dirichlet ensemble, $\alpha$ is boundary data: varying it is not
a change of state of the black hole, but an operation performed from the AdS boundary that changes the asymptotic tilt of the preferred frame. The term
proportional to $\beta\,\delta\alpha$ in Eq.~\eqref{eq:Dirichlet-thermal-first-law} is the corresponding work, with the response $(3c_{123}/2G)\beta$ acting as the
generalized force conjugate to $\alpha$. The Legendre transformation to $M_{\rm N}$ is then the analogue of the passage from internal energy to
enthalpy, and the Neumann work term in Eq.~\eqref{eq:Neumann-thermal-first-law} is the cost of changing the response at fixed misalignment. As for the
pressure--volume term of extended black-hole thermodynamics, both terms arise from varying data that the Hamiltonian treats as fixed, and neither affects
the horizon heat. Moreover, since the {\ae}ther symmetry of Ref.~\cite{ArataEtAl2026Elliptic} shifts $\beta$ at a fixed $\alpha$, as can be seen from Eq.~\eqref{eq:alpha-beta-transformation}, in the Neumann ensemble this symmetry acts as a work-producing deformation, with a work proportional to $\alpha$, just like the elliptic flux. In the aligned sector, which is thus singled out as the zero-source Dirichlet phase space, both work terms vanish and the first law closes without boundary contributions.

The second lesson concerns the thermodynamic reading of mass formul\ae. Written in terms of horizon quantities, the variation of $M_{\rm K}$ mixes horizon heat
and boundary work, so the coefficient of $\delta A$ need not reproduce the peeling temperature. This mismatch signals neither a failure of the tunnelling
temperature nor an inconsistency of the first law: once the phase space is specified, the heat is local to the universal horizon and governed by
$T_{\textsc{uh}}$, while the misalignment contributes only through work at infinity. More generally, in the presence of a dynamical preferred frame, the
energy entering the first law depends not only on the solution, but also on the boundary conditions imposed on the frame at infinity.

Our analysis is restricted to the static, spherically symmetric $c_{14}=0$ sector. Extending it to $c_{14}\neq0$ and to rotating solutions, where
Eq.~\eqref{eq:four-dimensional-counterterm} no longer fixes a unique covariant counterterm, requires the full boundary variational problem of Sec.~\ref{sec:CPS_boundaries} and, above all, a better understanding of the quasi-universal horizons that so far appear to be associated with rotating black holes in Einstein--{\ae}ther
theory~\cite{Franzin:2023rdl,DelPorro:2025zyv}. It would also be interesting to enlarge the phase space by boundary degrees of freedom whose symplectic form
cancels Eq.~\eqref{eq:integrability-obstruction}, so that both asymptotic modes could vary without obstruction.

We plan to return to these questions in the near future. More broadly, we hope that this work may help others appreciate the far-from-trivial lessons that
theories without local Lorentz invariance can teach about the inner workings of black-hole thermodynamics, whatever one's view of their ultimate role as a description of Nature. By breaking a fundamental local symmetry of spacetime, these theories lift degeneracies that general relativity keeps hidden, and, in
doing so, they expose what black-hole thermodynamics actually rests upon. Here, the same arbitrarily fast modes that make the universal horizon a causal boundary also tie it to the asymptotic region through a direct boundary-to-boundary holographic link, which works both ways: regularity at the universal horizon fixes the boundary response $\beta$ in terms of $r_\textsc{uh}$, Eq.~\eqref{eq:beta-regularity}, while the boundary conditions imposed on the preferred frame at infinity decide whether the horizon charge is integrable. Sometimes, it is only by letting go of one's most cherished assumptions that one comes to see the true essence of things.

\acknowledgments

GN is supported by the 1st edition (2025) of ``Research fellowship in fundamental physics and study of the universe'' established by the Antonio Madonna Foundation ETS. The authors used ChatGPT (OpenAI) and Claude (Anthropic) as assistive tools for discussing aspects of the physical results, checking proofs, and polishing the manuscript. All AI-assisted suggestions were critically assessed, independently checked, and approved by the authors, who take full responsibility
for the final content.

\appendix

\section{Generalized Komar normalization}
\label{app:Komar}

Let $\omega^{\mu\nu}=-\omega^{\nu\mu}$ denote a Killing potential satisfying
\begin{equation}
  \nabla_\mu\omega^{\mu\nu}=\chi^\nu.
  \label{eq:Killing-potential}
\end{equation}
Let $\dd S_{\mu\nu}$ denote the oriented surface element on $\partial\Sigma$ and $\dd\Sigma^\nu$ the directed hypersurface element on $\Sigma$.  Combining the Killing identity with the Einstein equation gives
\begin{equation}
  \begin{aligned}
    \frac{1}{8\pi G}\int_{\partial\Sigma}
      \bigl(\nabla^\mu\chi^\nu+\Lambda\omega^{\mu\nu}\bigr)\,\dd S_{\mu\nu} +\frac{1}{4\pi G}\int_\Sigma
      \left(T^{\aet}_{\mu\nu}-\frac12g_{\mu\nu}T^{\aet}\right)
      \chi^\mu\,\dd\Sigma^\nu=0.
  \end{aligned}
  \label{eq:generalized-Komar}
\end{equation}
The exterior hypersurface extends from the universal horizon to infinity.  In the coordinates of Eq.~\eqref{eq:EF-metric}, we choose the spherical
representative $\omega^{rv}=r/3$ and fix its homogeneous ambiguity by
matching to the maximally symmetric reference background
\cite{KastorRayTraschen2009,MartinMattingly2024}.

The metric Komar integral on a sphere of radius $r$ is
\begin{equation}
  \frac{r^2e'}{2G}
  =\frac{1}{G}\left(
    -\frac{\Lambda r^3}{3}+\frac{r_0}{2}
    +\frac{2c_{13}\beta^2}{r^3}
  \right).
  \label{eq:metric-Komar-sphere}
\end{equation}
The Killing-potential contribution cancels the cosmological divergence.  The
remaining \ae ther bulk integral and the universal-horizon surface term are,
respectively,
\begin{align}
  I_{\mathrm{bulk}}^{\aet}
  &=\frac{1}{G}\frac{c_{13}}{1-c_{13}}
    \left(\frac{2\ruh}{3}+\frac{2\ruh^3}{\ell_u^2}\right),
  \label{eq:Komar-aether-bulk}\\
  I_{\textsc{uh}}
  &=\frac{1}{G}\left[
    \frac{2\ruh}{3(1-c_{13})}
    +\frac{2\ruh^3}{(1-c_{13})\ell_u^2}
    -\frac{\ruh^3}{\ell_u^2}+\alpha\beta
  \right].
  \label{eq:Komar-horizon-term}
\end{align}
Taking their difference reproduces Eq.~\eqref{eq:Komar-mass},
\begin{equation}
  I_{\textsc{uh}}-I_{\mathrm{bulk}}^{\aet}=\MK.
  \label{eq:Komar-final}
\end{equation}
A comparison of
Eqs.~\eqref{eq:metric-Komar-sphere} and \eqref{eq:metric-charge} also confirms
the factor of two between the metric Komar integral and the metric Iyer--Wald
charge.

\section{Calculating the symplectic potential for the static spherically symmetric \texorpdfstring{$c_{14}=0$}{c14 = 0} case}
\label{app:symplectic-potential}
In this appendix, we show how to go from Eq.~\eqref{eq:aether-potential-boundary} to Eq.~\eqref{eq:aether-potential-sphere}.
As in Ref.~\cite{ArataLiberatiNeri2026}, we start from Eq.~\eqref{eq:aether-potential-boundary} and denote its three bracketed terms by $\mathcal B_1$, $\mathcal B_2$, and $\mathcal B_3$, in order.  The Latin indices $i,j,k$ refer to directions tangent to the boundary, while the Greek indices are four-dimensional.

The boundary variations are those in Eq.~\eqref{eq:cutoff-variations}.  Since $n_\mu\sim\dd r$, we do not need $\delta u_r$, because the tangential part of $\delta u$ is simply $\delta u_t$.

Using the definition of $Y^\mu{}_\rho$ in Eq.~\eqref{eq:Y-tensor}, the four components that enter our computation are
\begin{subequations}
  \label{eq:app-Y-components}
  \begin{align}
    Y^{tt}
    ={}&\frac{1}{e^2}\Bigl[-c_{13}(s\cdot\chi)^2(s\cdot\chi)'
      +c_2e\left(\frac{2(s\cdot\chi)}{r}+(s\cdot\chi)'\right)
      +c_3(s\cdot\chi)(u\cdot\chi)(u\cdot\chi)'\Bigr],
      \label{eq:app-Ytt}\\
    Y^{tr}
    ={}&\frac{1}{e}\Bigl[-c_{13}(u\cdot\chi)(s\cdot\chi)(s\cdot\chi)'
      +c_3(s\cdot\chi)^2(u\cdot\chi)'\Bigr],
      \label{eq:app-Ytr}\\
    Y^{rt}
    ={}&\frac{1}{e}\Bigl[-c_{13}(u\cdot\chi)(s\cdot\chi)(s\cdot\chi)'
      +c_3(u\cdot\chi)^2(u\cdot\chi)'\Bigr],
      \label{eq:app-Yrt}\\
    Y^{rr}
    ={}&-c_{13}(u\cdot\chi)^2(s\cdot\chi)'
      -c_2e\left(\frac{2(s\cdot\chi)}{r}+(s\cdot\chi)'\right)
      +c_3(s\cdot\chi)(u\cdot\chi)(u\cdot\chi)'.
      \label{eq:app-Yrr}
  \end{align}
\end{subequations}
\paragraph{\texorpdfstring{$\mathcal B_1$}{B1} Term:}

\begin{align}
  \mathcal B_1
  &\coloneqq n_\mu\bigl(u^\mu Y^{ij}-u^iY^{\mu j}-u^iY^{j\mu}\bigr)\delta q_{ij}
    \notag\\
  &=n_r\bigl(u^rY^{tt}-u^tY^{rt}-u^tY^{tr}\bigr)\delta q_{tt}
    \notag\\[0.4ex]
  &=\frac{-\delta e}{\sqrt e}\Biggl\{
      -\frac{(s\cdot\chi)}{e^2}\Biggl[
         -c_{13}(s\cdot\chi)^2(s\cdot\chi)' +c_2e\left(\frac{2(s\cdot\chi)}{r}+(s\cdot\chi)'\right)
    \notag\\*
  &\hspace{23em}+\Ccancel[red]{c_3(s\cdot\chi)(u\cdot\chi)(u\cdot\chi)'}
      \Biggr]
    \notag\\
  &\hspace{3.4em}+\frac{(u\cdot\chi)}{e^2}\Bigl[
        -c_{13}(u\cdot\chi)(s\cdot\chi)(s\cdot\chi)' +c_3(u\cdot\chi)^2(u\cdot\chi)'\Bigr]
    \notag\\
  &\hspace{3.4em}+\frac{(u\cdot\chi)}{e^2}\Bigl[
        -c_{13}(u\cdot\chi)(s\cdot\chi)(s\cdot\chi)' +\Ccancel[red]{c_3(s\cdot\chi)^2(u\cdot\chi)'}\Bigr]
      \Biggr\}
    \notag\\[0.4ex]
  &=\frac{\delta e}{e^2\sqrt e}\Biggl[
      -c_{13}(s\cdot\chi)^3(s\cdot\chi)'
      +c_2(s\cdot\chi)e\left(\frac{2(s\cdot\chi)}{r}+(s\cdot\chi)'\right)
    \notag\\*
  &\hspace{12.2em}-c_3(u\cdot\chi)^3(u\cdot\chi)'
      +2c_{13}(u\cdot\chi)^2(s\cdot\chi)(s\cdot\chi)'\Biggr].
  \label{eq:app-B1}
\end{align}
\paragraph{\texorpdfstring{$\mathcal B_2$}{B2} Term:}
\begin{equation}
  \begin{aligned}
    \mathcal B_2
    &\coloneqq2n_\mu Y^{\mu i}\delta u_i\\
    &=2n_rY^{rt}\delta u_t \\
    &=\frac{2\,\delta(u\cdot\chi)}{e\sqrt e}\Bigl[
      c_3(u\cdot\chi)^2(u\cdot\chi)'
      -c_{13}(u\cdot\chi)(s\cdot\chi)(s\cdot\chi)'\Bigr].
  \end{aligned}
  \label{eq:app-B2}
\end{equation}
\paragraph{\texorpdfstring{$\mathcal B_3$}{B3} Term:}
\begin{align}
  \mathcal B_3
  &\coloneqq2n_\mu n_\nu Y^{\mu\nu}\delta(n\cdot u)\notag\\
  &=2n_rn_rY^{rr}\delta(n\cdot u)
    \notag\\
  &=\frac{2}{e\sqrt e}\Biggl[
      c_{13}(u\cdot\chi)^2(s\cdot\chi)'
      +c_2e\left(\frac{2(s\cdot\chi)}{r}+(s\cdot\chi)'\right)
    \notag\\*
  &\hspace{11.1em}-c_3(s\cdot\chi)(u\cdot\chi)(u\cdot\chi)'\Biggr]
     \left(\delta(s\cdot\chi)-\frac{(s\cdot\chi)}{2e}\,\delta e\right)
    \notag\\
  &=\frac{2\,\delta(s\cdot\chi)}{e\sqrt e}\Biggl[
      c_{13}(u\cdot\chi)^2(s\cdot\chi)'
      +c_2e\left(\frac{2(s\cdot\chi)}{r}+(s\cdot\chi)'\right) -c_3(s\cdot\chi)(u\cdot\chi)(u\cdot\chi)'\Biggr]
    \notag\\
  &\quad-\frac{(s\cdot\chi)\,\delta e}{e^2\sqrt e}\Biggl[
      c_{13}(u\cdot\chi)^2(s\cdot\chi)'
      +c_2e\left(\frac{2(s\cdot\chi)}{r}+(s\cdot\chi)'\right) -c_3(s\cdot\chi)(u\cdot\chi)(u\cdot\chi)'\Biggr].
  \label{eq:app-B3}
\end{align}
Summing the three contributions and simplifying, we obtain
\begin{align}
  &\mathcal B_1+\mathcal B_2+\mathcal B_3\notag\\
  &=\frac{1}{e\sqrt e}\Bigl[c_{13}(s\cdot\chi)(s\cdot\chi)'
      -c_3(u\cdot\chi)(u\cdot\chi)'\Bigr]\delta e
    \notag\\
  &+\frac{2\,\delta(u\cdot\chi)}{e\sqrt e}\Bigl[
      c_3(u\cdot\chi)^2(u\cdot\chi)'
      -c_{13}(u\cdot\chi)(s\cdot\chi)(s\cdot\chi)'\Bigr]
    \notag\\
  &+\frac{2\,\delta(s\cdot\chi)}{e\sqrt e}\Biggl[
      c_{13}(u\cdot\chi)^2(s\cdot\chi)'
      +c_2e\left(\frac{2(s\cdot\chi)}{r}+(s\cdot\chi)'\right)
      -c_3(s\cdot\chi)(u\cdot\chi)(u\cdot\chi)'\Biggr].
  \label{eq:app-B-sum}
\end{align}
Using the fact that $e=(u\cdot\chi)^2-(s\cdot\chi)^2$, we get the relation among their variations
\begin{equation}
  \delta e=2(u\cdot\chi)\,\delta(u\cdot\chi)
    -2(s\cdot\chi)\,\delta(s\cdot\chi).
  \label{eq:app-norm-variation}
\end{equation}
Thus,
\begin{align}
  &\mathcal B_1+\mathcal B_2+\mathcal B_3 \notag\\
  &=\frac{1}{e\sqrt e}\Bigl[c_{13}(s\cdot\chi)(s\cdot\chi)'
      -c_3(u\cdot\chi)(u\cdot\chi)'\Bigr]
      \Bigl[\delta e-2(u\cdot\chi)\,\delta(u\cdot\chi)\Bigr]
    \notag\\
  &+\frac{2\,\delta(s\cdot\chi)}{e\sqrt e}\Biggl[
      c_{13}(u\cdot\chi)^2(s\cdot\chi)'
      +c_2e\left(\frac{2(s\cdot\chi)}{r}+(s\cdot\chi)'\right)
      -c_3(s\cdot\chi)(u\cdot\chi)(u\cdot\chi)'\Biggr]
    \notag\\
  &=\frac{2\,\delta(s\cdot\chi)}{e\sqrt e}\Biggl[
      c_{13}\bigl((u\cdot\chi)^2-(s\cdot\chi)^2\bigr)(s\cdot\chi)'
      +c_2e\left(\frac{2(s\cdot\chi)}{r}+(s\cdot\chi)'\right)\Biggr]
    \notag\\
  &=\frac{2}{\sqrt e}\left[c_{123}(s\cdot\chi)'
    +\frac{2c_2}{r}(s\cdot\chi)\right]\delta(s\cdot\chi).
  \label{eq:app-B-final}
\end{align}
Multiplying by the boundary volume element and the factor
$-1/(16\pi G)$, contracting with $\chi$, and integrating
over the two-sphere yields Eq.~\eqref{eq:aether-potential-sphere}.
Here, the factor $1/\sqrt{e}$ cancels against
$\chi\cdot\boldsymbol\epsilon_\Gamma
=-\sqrt{e}\,\boldsymbol\epsilon_{S^\infty_r}$,
and the angular integration contributes $4\pi$.

\bibliographystyle{JHEP}
\bibliography{references}

\end{document}